\documentclass[
reprint,
nofootinbib,
amsmath,amssymb,amsbsy,
prd,
twocolumn,
superscriptaddress,
longbibliography,
preprintnumbers]{revtex4-2}
 
\usepackage[T1]{fontenc}
\usepackage[shortlabels]{enumitem}
\usepackage{pifont}
\usepackage{amsmath}
\usepackage{xcolor}
\usepackage[mathscr]{euscript}
\usepackage{bm}
\newcommand{\mybar}[1]%
        {\kern 0.6pt\overline{\kern -0.6pt#1\kern -0.6pt}\kern 0.6pt}
\makeatletter
\newcommand{\subalign}[1]{%
  \vcenter{%
    \Let@ \restore@math@cr \default@tag
    \baselineskip\fontdimen10 \scriptfont\tw@
    \advance\baselineskip\fontdimen12 \scriptfont\tw@
    \lineskip\thr@@\fontdimen8 \scriptfont\thr@@
    \lineskiplimit\lineskip
    \ialign{\hfil$\m@th\scriptstyle##$&$\m@th\scriptstyle{}##$\hfil\crcr
      #1\crcr
    }%
  }%
}
\makeatother

 \usepackage{mathtools}
 \usepackage{lmodern}
 \usepackage{lmodern}
\usepackage{graphicx}
\usepackage{hyperref}
\usepackage{amsfonts}
\usepackage{tocvsec2}
\usepackage{slashed}
\usepackage{cancel}
\usepackage{comment}

\newcommand\beq{\begin{eqnarray}}
\newcommand\eeq{\end{eqnarray}}

\newcommand\eqn[1]{\label{eq:#1}}

\newcommand\Eq[1]{Eq.~(\ref{eq:#1})}

\def\U\Omega{U(1)_{\Omega}}

\hypersetup{
  colorlinks=true,   
  linkcolor=cyan,    
  citecolor=blue,    
  filecolor=magenta, 
  urlcolor=blue      
}
\usepackage[section]{placeins}

\begin{document}
\title{Ubiquity of ocular resonances in scattering phenomena}

\author{Cullen Gantenberg}
\email{cgantenb@uw.edu}
\affiliation{Institute for Nuclear Theory, Box 351550, Seattle, Washington 98195-1550}
\author{David B. Kaplan}
\email{dbkaplan@uw.edu}
\affiliation{Institute for Nuclear Theory, Box 351550, Seattle, Washington 98195-1550}
\preprint{INT-PUB-26-035}

\begin{abstract}
We demonstrate how quantum mechanical scattering off a potential with a characteristic surface generically leads to a family of resonances in the complex momentum plane clustered near a roughly eye-shaped curve centered on the origin and formed by anti-Stokes lines.  A particularly interesting class are  potentials that are more singular than $1/r^2$ at short distance, where the scattering surface is the separatrix between captured and deflected classical trajectories. These resonances have been recently shown to play an important role in low energy nucleon scattering, and can be related to ``creeping waves'' familiar  from classical diffractive scattering and analyzed by means of Regge poles.
\end{abstract}

\maketitle

\section{Introduction}
An important feature in many quantum mechanical scattering processes is the existence of resonances. In this Letter we argue that there is a class of low-lying resonances that possess universal features and are found in a wide class of non-relativistic scattering problems, examples of which were recently found to play a surprisingly important role in low energy nucleon-nucleon scattering \cite{gantenberg2026dimer}. These resonances are poles in the scattering amplitude clustered near anti-Stokes lines that form a roughly eye-shaped curve  about the origin in the complex momentum plane, and we therefore dub them ``ocular resonances''.  Such resonances are related to a wave phenomenon, the so-called creeping wave, which plays a role in the description of the optical phenomenon known as a glory, or in diffractive scattering by black holes \cite{nussenzveig1963mandelstam,nussenzveig1991complex,nussenzveig1992diffraction,folacci2019regge}. We find that they not only occur when scattering off hard surfaces, such as one normally associates with diffraction, but also from scattering by purely attractive  potentials more singular than $1/r^2$ at short distance, where the role of the scattering surface is played by the sphere of unstable classical orbits, at the radius where the centrifugal and central forces cancel.  Such potentials  are ubiquitous in nuclear and atomic physics. We use the WKB approximation to derive the existence of such resonances in the large-$\ell$ limit, where $\ell$ is the angular momentum of a particular partial wave, and then show that the semiclassical description remains  quantitatively correct even at low angular momentum.

 \section{Semiclassical Techniques}

 We begin by considering scattering from  a central potential $V(r)$ in the limit  of large angular-momentum  $\ell$.
Introducing rescaled variables
\begin{equation}
r=L^p\rho,\quad
\kappa=L^{p-1}k,\quad
\Omega_L(\rho)=L^{2p-2}M V(L^p\rho) \,,
\eqn{rescaled}
\end{equation}
where $L=(\ell+1/2)$, the reduced radial Schr\"odinger equation becomes
\begin{equation}
\left[
\frac{1}{L^2}\frac{d^2}{d\rho^2}
+\kappa^2-\frac{1}{\rho^2}+\frac{1}{4L^2\rho^2}
-\Omega_L(\rho)
\right]\psi(\rho)=0\ .
\eqn{Seq}
\end{equation}
  A nontrivial semiclassical limit exists 
when the scaling power $p$ and  the parameters of the potential can be chosen such that 
$
\Omega_L(\rho)\to\Omega_\infty(\rho)
$ 
with $\Omega_\infty$ finite and nonzero as $L\to\infty$.

\begin{figure}[t]
     \centering
     \includegraphics[width=0.75\linewidth]{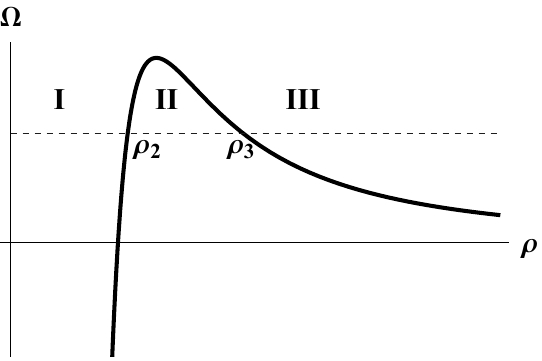}
     \caption{The solid black curve is the schematic effective potential for an attractive singular potential with turning points $\rho_2$ and $\rho_3$  marked for a particular energy $\kappa^2$ (dashed).  }
     \label{fig:effpot}
 \end{figure}

For example, for a singular power-law potential $V(r) = g/r^N$ with $N>2$, rescaling by $p=\frac{2}{2-N}$
gives the finite large-$L$ limit
\beq
\Omega_\infty(\rho)=\frac{Mg}{\rho^N}\ .
\eqn{Om}
\eeq
If $V$ is not a power-law potential but has leading singularity $r^{-N}$ for small $r$, then upon the above rescaling one obtains the same limiting behavior as in \Eq{Om} -- the subleading singularities are ``irrelevant'' in a reversal of Wilsonian language.  For example, if $V(r)\propto \exp(-m r)/r^3$ one again finds $\Omega_\infty(\rho)\sim 1/\rho^3$.  This behavior is simply explained by the fact that the peak in Fig.~\ref{fig:effpot} moves to shorter distance with increasing $L$, where the exponential tail of the potential becomes unimportant.

A fixed finite-range potential behaves differently.  Suppose, for example, that
\begin{equation}
V(r)=g v(r/a)\,,
\end{equation}
where $a$ and $g$ are independent of $L$.  Keeping its physical range fixed requires scaling dimension $p=0$.  With
\begin{equation}
x=\frac{r}{a}\ ,\quad
\kappa=\frac{ka}{L}\ ,\quad \lambda_L\equiv\frac{Mga^2}{L^2}\ ,
\end{equation}
the potential $V(r)$ gets replaced by $\lambda_L v(x)$, and 
since
$
\lambda_L\to 0
$ as $L\to\infty
$,
the  finite-range interaction flows to the free theory at fixed $\kappa$.  This observation does not preclude considering hard-sphere scattering and its ilk so long as the $|g|\to\infty$ limit is taken before the large-$L$ limit.

Including the angular momentum barrier, scattering in the large-$L$ limit is determined by the semiclassical effective potential 
\begin{equation}
\Omega_{\rm eff}(\rho)
\equiv \frac{1}{\rho^2}+\Omega_\infty(\rho)\ .
\end{equation}
For potentials that are attractive and more singular than $1/r^2$  there are generically two real turning points\footnote{ In all problems analyzed here there are at most three independent turning points, labeled such that for small, real $\kappa$, $\mathrm{Re}\,\rho_1\leq 0<\rho_2<\rho_3$.} and hence three qualitatively distinct regions, as shown in Fig.~\ref{fig:effpot}. Analogous complex turning points may be defined when the potential is repulsive.  Scattering observables are determined by the relative coefficients of incoming and outgoing waves in the exterior region,
\begin{equation}
\psi_{III}(\rho)=
\frac{1}{\sqrt{\xi(\rho)}}
\left[
c_+e^{iL\Phi(\rho_3,\rho)}
+c_-e^{-iL\Phi(\rho_3,\rho)}
\right],
\end{equation}
where $\rho_3$ denotes the outer turning point and
\begin{equation}
\Phi(\rho_i,\rho_j)
=\int_{\rho_i}^{\rho_j}\xi(\rho')d\rho',
\quad
\xi(\rho)=
\sqrt{\kappa^2-\Omega_{\rm eff}(\rho)}.
\eqn{xidef}
\end{equation}
To ease notation, we will often denote the actions between turning points---or between a turning point and the origin--- $\Phi(\rho_i,\rho_j)$ as $\Phi_{ij}(\kappa)$, making explicit the dependence on $\kappa$. At leading semiclassical order, matching the solution specified by its ultraviolet (UV) boundary condition at the origin, across the inner turning point $\rho_2$, and subsequently across $\rho_3$ gives, 
\begin{equation}
\begin{pmatrix}
c_+\\[2pt] c_-
\end{pmatrix}
=
{\bf M}
\begin{pmatrix}
e^{+iL\Phi_{23}(\kappa)}\\[2pt]
e^{-iL\Phi_{23}(\kappa)}
\end{pmatrix},
\eqn{cpm}
\end{equation}
where the  matrix ${\bf M}$ encodes the turning-point connection formulae and the UV boundary condition. 
An explicit calculation for an effective potential as in Fig.~\ref{fig:effpot} reveals that 
\begin{equation}
    c_-=e^{i\pi/4}\left( \sin\delta(\kappa) e^{iL\Phi_{23}(\kappa)}-\frac{i}{2}e^{i\delta(\kappa)}e^{-iL\Phi_{23}(\kappa)}\right)
    \eqn{cminus}
\end{equation} 
where
    $\delta(\kappa)\equiv\gamma+\pi/4- L\Phi_{02}(\kappa)$, and we define $\Phi_{0i}\equiv\lim_{\epsilon\to 0}\left(\Phi(\epsilon,\rho_i)-1/\epsilon \text{ terms}\right)$, the singularity removed from the action  with the real phase $\gamma$  defining the UV boundary condition.  The derivation of these relations and their validity in the $\kappa$ plane are discussed in Appendix~\ref{app:cminus}. With these definitions the resonance condition $c_-=0$ becomes 
\begin{equation} e^{2i\delta(\kappa)}\left(1+e^{-2iL\Phi_{23}(\kappa)}\right)=1.
\eqn{Rescond}
\end{equation}
In the cases we consider, $\mathrm{sgn}( \mathrm{Im}\,\delta)=-\mathrm{sgn}(\mathrm{Im}\,\kappa^2)$, so in order to have a solution at large $L$ we must demand that the two exponentials balance in quadrant four, while the term in parentheses must be exponentially small in quadrant three. This leads to the conditions
\begin{equation}
\begin{aligned}
 \mathrm{Im}\,\Phi_{23}(\kappa)=0 ,\quad &\mathrm{Re\,\kappa<0},\\
   \mathrm{Im}\,\Phi_{03}(\kappa)=0 ,\quad &\mathrm{Re\,\kappa>0},
\end{aligned}
\eqn{master}
\end{equation}
\begin{figure}[t]
   \centering
    \includegraphics[width=.8\linewidth]{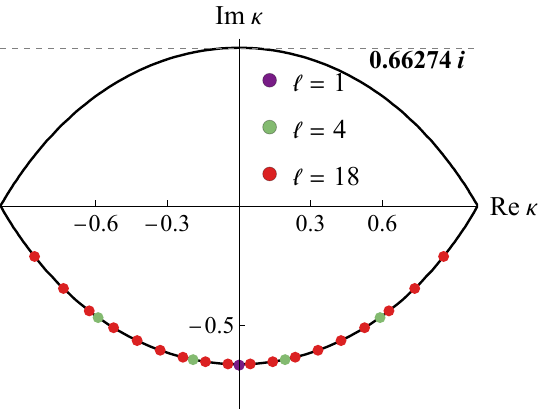}
    \caption{
      Resonance poles for scattering from an infinitely deep square well, plotted in the complex $\kappa$-plane for $\ell=1,4,18$. Each partial wave exhibits $\ell$ zeros sitting almost exactly on the semiclassical Stokes eye (black curve), even for $\ell=1$.
    }
    \label{fig:eye}
 \end{figure}

\noindent
using $  \left[\Phi_{02}(\kappa)+\Phi_{23}(\kappa)\right]=\Phi_{03}(\kappa)$ in the second line. The turning points $\rho_i(\kappa)$ and relevant integrals are to be analytically continued into the complex $\kappa$ plane, with the appropriate choice of Riemann sheet. For the class of attractive and repulsive interactions studied below, \Eq{master} defines a closed contour surrounding the origin in the complex $\kappa$ plane, comprised of anti-Stokes lines across which terms exponentially growing and decaying in $L$ are exchanged.  The contour has a characteristic ``eye'' shape whose size is determined by the momenta of the unstable orbit at the top of the effective potential.  For more complicated turning point geometries such that \Eq{cminus} fails to hold in the entire lower half plane, we still observe closed curves defined by the stitching together of various action curves, the network of which we refer to as the \emph{Stokes eye}. This eye traces the limiting locus of ocular resonance poles as $L\to\infty$.  Deviation of poles from the Stokes eye scales as $O(1/L^2)$ except at stitching points. There will be $O(L)$ resonance poles whose location is determined by phase cancellation between terms in \Eq{cminus}, which depends on the short-distance boundary condition through $\delta$; changing that boundary condition shifts the poles along the contour of the eye and, at exceptional values, may produce finely tuned resonances, bound, or virtual states within the eye. 

\subsection{Infinite square well scattering}

 Scattering from an infinitely deep spherical square well of radius $a=1$ gives the simplest example of ocular resonances.  The  semiclassical behavior is the same whether the square well is infinitely attractive or infinitely repulsive as the two cases have identical $S$-matrices, up to fine-tuned points in parameter space where a bound state sits at threshold,
despite the wave functions in the two interiors being qualitatively very different.

The appropriate scaling power is $p=0$  with
$
    \rho=r$, 
   $ \kappa=k/L
$.
The potential vanishes for $\rho>1$, and the reduced radial Schr\"odinger equation in the exterior region at leading order is therefore
\begin{equation}
\left[
\frac{1}{L^2}\frac{d^2}{d\rho^2}
+\kappa^2-\frac{1}{\rho^2}
\right]\psi(\rho)=0 ,
\qquad \rho>1 .
\end{equation}

For real $\kappa<1$, the centrifugal barrier has a turning point at
$
    \rho_3=1/\kappa
$,
while the  boundary at $\rho_2=1$ provides the inner reflection point. The interior region decouples in the infinite well limit and the relevant WKB action is
\beq
    \Phi_{23}(\kappa)
    &=&
    \int_1^{1/\kappa}
    d\rho\,
    \sqrt{\kappa^2-\frac{1}{\rho^2}} \cr
&=&
    -\sqrt{\kappa^2-1}
    +
    \tan^{-1}\left(\sqrt{\kappa^2-1}\right)\ .
\eeq
The condition that the two semiclassical contributions have equal magnitude is therefore
\begin{equation}
    \operatorname{Im}
    \left[
    -\sqrt{\kappa^2-1}
    +
    \tan^{-1}\left(\sqrt{\kappa^2-1}\right)
    \right]
    =0 .
\label{eq:bessel-eye}
\end{equation}
The Stokes eye defined by \Eq{bessel-eye} is shown in Fig.~\ref{fig:eye}.  

In this case the semiclassical result may be compared directly with the
exact solution.  The partial-wave $S$ matrix is given by
$
    S_\ell(k)
    =
    -
   {h_\ell^{(2)}(k)}/
         {h_\ell^{(1)}(k)}
$,
so the resonance poles are at the zeros of the  Hankel function,
\begin{equation}
    h_\ell^{(1)}(k)= h_\ell^{(1)}(L\kappa)
    =
    \sqrt{\frac{\pi}{2L\kappa}}\,
    H_L^{(1)}(L\kappa)=0\ .
\end{equation}
The semiclassical large-$L$ limit is then equivalent to a uniform
 asymptotic expansion of the Hankel function in the relevant
Stokes region, see Refs. \cite{Olver1954Bessel,Olver1997Asymptotics}.
Fig.~\ref{fig:eye} shows that even for small $\ell$ the exact zeros
lie remarkably close to the Stokes eye.  For example, the lower edge of the limiting
curve intersects the imaginary axis at the Laplace limit,
$
    \kappa=-i(0.662743\ldots)
$ while at $\ell=1$ the exact spherical
Hankel zero is instead
 $ \kappa=-{2i/3}$,
a discrepancy of  $0.6\%$.  

Along the Stokes eye, the two semiclassical exponentials have
equal moduli and the resonance condition $c_-=0$ is determined by the
relative phase $L\,\operatorname{Re}\Phi_{23}(\kappa)$, which predicts $\ell$ resonances, 
in agreement with the exact result.

The infinite spherical square well provides a useful description for more general
potentials.  Whenever an interaction produces sufficiently sharp and nearly
complete reflection from a narrow range of radii, the exterior wave can
behave approximately as though it were reflected from a hard surface;
the corresponding resonance poles will then lie close to a rescaled version of this same Stokes
eye defined by the Hankel function, with    phase and radius determined by the
underlying interaction. We conjecture that this same curve is where resonance poles lie for attractive singular potentials $\Omega\sim-\rho^{-N}$ in the large-$N$ limit, although we show below that the behavior is more complicated for finite $N$\footnote{The correspondence between the hard sphere and the repulsive singular potentials in the large $N$ limit has been studied via Regge pole techniques by \cite{Connor}.}.

\subsection{Attractive Singular Potentials}

We now turn to the analogous structures of attractive singular potentials, which  arise naturally in nuclear and atomic physics, all of which can be put into the form
\begin{equation}
    \Omega_{\infty}(\rho)=-1/\rho^N,
\end{equation}
for $N>2$, with the coupling rescaled into the definition of the coordinates and momentum. 

\begin{figure*}[t]
    \centering
    \includegraphics[width=1.0\linewidth]{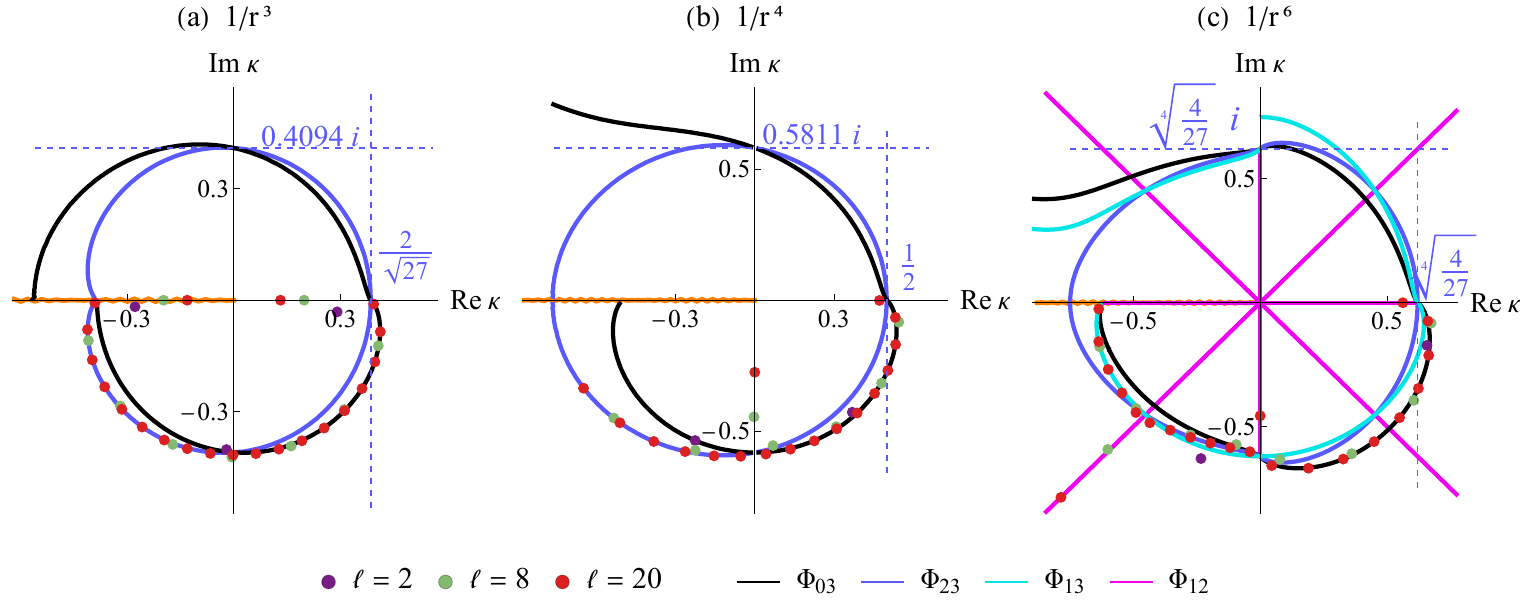}
    \caption{The semiclassical Stokes eye determined analytically for the indicated cycles for the attractive $1/r^3$, $1/r^4$ and $1/r^6$ potentials. Superimposed are resonance pole locations for  angular momenta $\ell=2,8,20$, with UV boundary condition $\gamma=1$, all lying  on the principal sheet in the complex $\kappa$ plane, with branch cut  taken along the negative real axis. Only the relevant cycles for each potential are shown. The repulsive cases yield qualitatively similar plots, although in this case the poles lie one sheet above or below  the principal sheet.}
    \label{fig:Peaches}
\end{figure*} 

For $N=3,4,6$ the barrier integrals can be computed exactly in terms of elliptic integrals which can be analytically continued to complex $\kappa$, allowing identification of the Stokes eye  once the branch cut is chosen\footnote{These  cases are all of physical interest: the tensor force in nucleon interactions and atomic dipole-dipole interactions are characterized by $N=3$; the interaction between an ion and a neutral atom scales as $N=4$; and the van der Waals potential between neutral atoms has $N=6$.}. One fundamental difference from the previous case is that for power law tails the partial wave $S$ matrix elements are now multi-sheeted in $\kappa$ instead of meromorphic. We will be concerned only with the relevant Stokes curves on the same Riemann sheet as the physical scattering axis.
As discussed above, the cases $N=3,4,6$ exhibit nontrivial large $L$ limits for scaling dimensions $p=-2,-1,-1/2$, respectively. The turning points $\rho_{2,3}$ merge at the dimensionless energy $\kappa_c^2$ corresponding to the unstable orbit at the top of the angular momentum barrier, $\kappa_c^2=4/27,1/4,\sqrt{4/27}$, respectively, and the Stokes curves will therefore intersect the positive real axis at $+\kappa_c$ as \Eq{master} is  trivially satisfied when $\rho_2=\rho_3$.
  After performing the analytic continuation of the action integrals, we identify the relevant geometries to be the Stokes eyes displayed in Fig.~\ref{fig:Peaches} -- less symmetric than Fig.~\ref{fig:eye} since time reversal can act nontrivially between Riemann sheets. In addition to the Stokes eyes for each potential we show resonance pole locations, computed numerically by evaluating the Jost function along a complex-rotated contour, for several angular momenta and all defined by the (arbitrary) UV boundary condition $\gamma=1$ at the origin. Fig.~\ref{fig:Peaches} also exhibits finely tuned barrier resonances close to the real axes, and virtual states on the lower imaginary axes for $N=4,6$. It is clear from the plots that \Eq{cminus} correctly describes the action switch in both the $1/r^3$ and $1/r^4$ potentials, occurring when the curves intersect at the lower imaginary axis. Moreover, the poles predicted by \Eq{Rescond} are accurate to within a few percent at low $\ell$. The $1/r^6$ potential is evidently richer in structure, requiring four unique cycles to describe the network of poles. The introduction of additional cycles results from a change in the topology of the Stokes graph formed by turning points and their connections by Stokes lines in the complex $\rho$ plane, which changes how one matches the wavefunction from the origin to the exterior region. We find  that ocular resonances also exist for repulsive power-law potentials, residing not on the principal sheet but on adjacent sheets, the relevant energy scale again given by the  unstable orbit of the corresponding attractive potential.

\section{Discussion} 
Ocular resonances  are quite different than the long-lived resonances corresponding to poles sitting close to the real axis,  interpreted as particles trapped temporarily behind a classical barrier. What are we to make of these clustered poles that are much farther from the real axis -- although  closer to the origin than  all but exceptional resonances?
At least in the case of scattering off hard objects the interpretation is known and is best revealed through the complex angular momentum picture, i.e. of complex $\ell$ eigenvalues of the radial Schr\"odinger operator. For the hard sphere, these Regge poles form a tower of states in the first and third quadrants of the complex $\ell$-plane. The contributions of these states to the scattering amplitude lend to their interpretation as waves that ``creep'' or diffract around hard surfaces \cite{nussenzveig1991complex,nussenzveig1992diffraction}. Such states are related to ocular resonances in the following sense:  the Regge pole closest to the positive real $\ell$-axis is responsible for the ocular state closest to the positive real $\kappa$-axis in every partial wave, with the next Regge pole being responsible for the next closest, and so on. In this way, we can interpret these ocular states as partial wave manifestations of diffraction. What is novel is that these states are not restricted to hard surface potentials as might be expected, but are seen in scattering from attractive singular potentials as well. Furthermore, while we have revealed these resonances using large-$\ell$ semiclassical techniques, it is apparent from our results that the resulting description is quantitatively accurate for low-$\ell$ as well, as was discovered for nucleon-nucleon scattering~\cite{gantenberg2026dimer}.

\begin{acknowledgments}
 This research is supported in part by Department of Energy Grant No. DE-FG02-00ER41132.  We thank J.J. Segura Sala for communication. ChatGPT 5.6 and Claude Opus 5.5 were used to aid code development and analysis of analytic structures.
\end{acknowledgments}

\appendix
\section{Computation of $c_-$}
\label{app:cminus}
\begin{figure}
    \centering
    \includegraphics[width=0.95\linewidth]{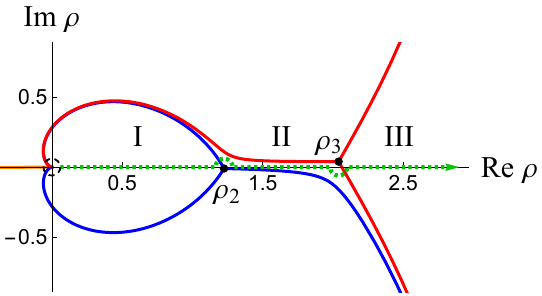}
    \caption{Typical turning point geometry associated with an effective potential as in Fig.~\ref{fig:effpot}, for $\kappa$ just below the positive real axis.  The integration path taken is the dashed green curve from $\rho=0^+$ to $\rho=+\infty$, avoiding the turning points where WKB solutions are ill-defined. Emanating from each turning point are three Stokes lines, the crossing of which alters the coefficients of subdominant waves in the WKB approximation. This particular geometry is for $1/r^3$, the leftmost turning point not shown.}
    \label{fig:tpgeom}
\end{figure}
In this section we derive \Eq{cminus}, constructing the asymptotic coefficients of the exterior waves for the relevant turning point geometry shown in Fig.~\ref{fig:tpgeom} plotted in the complex $\rho$ plane. 
A particular WKB solution is determined by the boundary condition at the origin, and is then transported along the dashed green line in the complex $\rho$ plane.  Embedded in the plane is a Stokes graph whose vertices are the turning points and the singular point at the origin; emanating from the vertices are Stokes lines, defined by $\mathrm{Re}\,\Phi(\rho_i,\rho)=0$ where $\rho_i$ is the vertex. These lines must end on another vertex or go to infinity. The Airy connection formulas (see \cite{Heading1962,Berry1972Semiclassical}) determine how the coefficients of subdominant exponentials---those that decay with increasing $L$---jump upon crossing these lines, as is required to ensure that $\psi(\rho)$ is smooth. The general rule is that the crossing of Stokes lines as in Fig.~\ref{fig:tpgeom} yields a change in the coefficient of the subdominant exponential equal to $\pm i$ times the coefficient of the dominant exponential, with the upper and lower sign corresponding to traversing the line in a counterclockwise or clockwise manner, respectively. Other connections are required when turning points merge or when the green path is made to sit on or arbitrarily close to a Stokes line without fully crossing it \cite{Berry1972Semiclassical}. We consider $\kappa$ to be below the real axis on the first sheet here; the analytic continuation to the entire lower half plane is valid so long as the  Stokes graph topology remains unchanged as $\kappa$ is rotated.  If the topology does change and a different pattern of Stokes lines is crossed by the transport contour, the solution must be recomputed.  

For the Stokes graph topology pictured in Fig.~\ref{fig:tpgeom} it is useful to represent the wavefunction in the three regions as
\begin{align}
\psi_{I}(\rho)&=
\frac{1}{\sqrt{\xi(\rho)}}
\left[
a_Ie^{iL\Phi(\rho,\rho_2)}
+b_Ie^{-iL\Phi(\rho,\rho_2)}
\right]\\
\psi_{II}(\rho)&=
\frac{e^{i\pi/4}}{\sqrt{\tilde{\xi}(\rho)}}
\left[
a_{II}e^{-L\tilde\Phi(\rho_2,\rho)}
+b_{II}e^{L\tilde\Phi(\rho_2,\rho)}
\right]\\
\psi_{III}(\rho)&=
\frac{1}{\sqrt{\xi(\rho)}}
\left[
c_+e^{iL\Phi(\rho_3,\rho)}
+c_-e^{-iL\Phi(\rho_3,\rho)}
\right]
\end{align}
where $\xi$ and $\Phi$ are defined in \Eq{xidef}, while in the tunneling region II we have $\tilde\xi=\sqrt{-\kappa^2+\Omega_{\text{eff}}(\rho)}$ and $\tilde\Phi(\rho_i,\rho_j)=\int_{\rho_i}^{\rho_j}\tilde\xi(\rho')d\rho'$. 
In order to match these solutions it is necessary to analytically continue the solution from one region to the next so that they can be compared. The appropriate analytic continuation follows from the orientation of the green contour and the behavior of $\xi$ and $\tilde\xi$ near the turning points. As the contour passes  above $\rho_2$, $\xi(\rho)\to-i\tilde{\xi}(\rho)$, so analytic continuation of the basis I into region II is given by
\begin{align}
        \frac{1}{\sqrt{{\xi}(\rho)}}
  \begin{pmatrix}
      e^{iL\Phi(\rho,\rho_2)}\\e^{-iL\Phi(\rho,\rho_2)}
  \end{pmatrix}
  \to
        \frac{e^{i\pi/4}}{\sqrt{\tilde{\xi}(\rho)}}
  \begin{pmatrix}
      e^{-L\tilde\Phi(\rho_2,\rho)}\\e^{L\tilde\Phi(\rho_2,\rho)}
  \end{pmatrix}.
\end{align}
Similarly, as the contour passes below $\rho_3$, $\tilde{\xi}(\rho)\to i\xi(\rho)$  and the continuation of the basis II into region III is given by 
\begin{align}
        \frac{e^{i\pi/4}}{\sqrt{\tilde{\xi}(\rho)}}
  \begin{pmatrix}
      e^{-L\tilde\Phi(\rho_2,\rho)}\\e^{L\tilde\Phi(\rho_2,\rho)}
  \end{pmatrix} \to
   \frac{1}{\sqrt{{\xi}(\rho)}}
  \begin{pmatrix}
      e^{-iL\Phi(\rho_3,\rho)}e^{-L\tilde\Phi(\rho_2,\rho_3)}\\e^{iL\Phi(\rho_3,\rho)}e^{L\tilde \Phi(\rho_2,\rho_3)}
  \end{pmatrix} .
\end{align}

The above relations allow us to transport the bases through the relevant turning points; relating the coefficients requires taking into account the crossings of Stokes lines near $\rho_2,\rho_3$. A straightforward calculation shows that upon crossing over the inner turning point $\rho_2$, the positive frequency wave is dominant. Using the above rules and the relevant continuation of basis I, it follows that  
\begin{align}
    a_I=a_{II}\ , \quad b_{II}=b_I-ia_I.
\end{align}
The chosen contour crosses no Stokes lines  in region II, so the solutions merely pick up the tunneling factors. Upon crossing the second Stokes line as the contour circumvents $\rho_3$, the positive frequency wave is again the dominant one. The continuation of basis II and the connection condition then give
\begin{align}
    c_+&=b_{II}e^{L\tilde\Phi(\rho_2,\rho_3)},\\
    c_-&=a_{II}e^{-L\tilde\Phi(\rho_2,\rho_3)}+ib_{II}e^{L\tilde\Phi(\rho_2,\rho_3)}.
\end{align}
The coefficient $c_-$ may therefore be expressed in $a_I$ and $b_I$ as
  \begin{align}
      c_-=a_Ie^{-L\tilde\Phi(\rho_2,\rho_3)}+i(b_I-ia_I)e^{L\tilde\Phi(\rho_2,\rho_3)}.
  \end{align}
For the $1/r^3$ potential, the coefficients of region I are determined by the boundary condition at $\rho=0^+$, given by  $\psi\sim \rho^{3/4}\cos(2L\sqrt{1/\rho}+\gamma)$. This fixes
\beq
    a_I=\frac{1}{2}e^{i\gamma-i\eta}\ ,\quad
     b_I=\frac{1}{2}e^{-i\gamma+i\eta}
\eeq
where 
\begin{equation}
    \eta\equiv\lim_{\epsilon\to0}\left(-2L/\sqrt{\epsilon} +L\Phi(\epsilon,\rho_2)\right),
\end{equation}
Defining   $\delta\equiv\gamma-\eta+\pi/4$ we obtain
\begin{equation}
    c_-=e^{i\pi/4}\left( \sin\delta e^{L\tilde\Phi(\rho_2,\rho_3)}-\frac{i}{2}e^{i\delta}e^{-L\tilde\Phi(\rho_2,\rho_3)}\right).
\end{equation}
The resonance condition $c_-=0$ then simplifies to 
\begin{equation}
    e^{2i\delta}\left(1+e^{-2L\tilde\Phi(\rho_2,\rho_3)}\right)=1,
\end{equation}
or, on replacing  $\tilde{\xi}(\rho)\to i\xi(\rho)$ and using the abbreviated notation for $\Phi$, 
\begin{equation}
    e^{2i\delta(\kappa)}\left(1+e^{-2iL\Phi_{23}(\kappa)}\right)=1,
\end{equation}
the result in \Eq{Rescond}. The generalization to other  potentials is straightforward.

\section{Stokes eyes for singular potentials}
\label{app:eyes}
Here we discuss how to construct the network of action curves $\mathrm{Im}\Phi_{ij}(\kappa)=0$ for the soluble singular potentials, using $1/r^3$ as an explicit example.
In this case, the barrier integral for the attractive $1/r^3$ potential may be written
\begin{equation}
\Phi(\rho_2,\rho_3)=
i\kappa\sqrt{\frac{\rho_2 - \rho_1}{\rho_3}}
\left[
(\rho_1-\rho_3)\,K\!\left(\Xi
\right)+
3\rho_3\,E\!\left(\Xi
\right)
\right],
\end{equation}
where 
\begin{equation}
    \Xi=\frac{\rho_1(\rho_3 - \rho_2)}
{\rho_3(\rho_1-\rho_2)}.
\end{equation}
The functions $K(\Xi)$ and $E(\Xi)$ are elliptic integrals of the first and second kinds, respectively, and $\rho_1,\rho_2,\rho_3$ are the roots of the cubic
\begin{equation}
    \kappa^2\rho^3-\rho+1=0,
\end{equation}
with $\rho_{2,3}$ designating the inner and outer turning points respectively for real, positive energies below the barrier top, as in Fig.~\ref{fig:effpot}. In the lower half plane, the other actions may be constructed by permuting the roots of the cubic into the appropriate slots; such a permutation does not define the other action integrals everywhere on the first sheet due to a multivaluedness of the representation, resulting from the branch structure of both the roots and the elliptic integrals. To simplify this analysis for $1/r^3$ we exploit the symmetry relation
\begin{equation}
\Phi_{12}(\kappa)+\Phi_{23}(\kappa)+\Phi_{31}(\kappa)=0,
\end{equation}
which defines the integration contours of these actions.
For example, to represent the action $\Phi_{12}(\kappa)$ in the upper half plane, we write $\Phi_{12}(\kappa)=\Phi_{13}(\kappa)-\Phi_{23}(\kappa)$ and use the Schwarz reflection principle to continue the function across the real axis. For functions $f(z)$ defined in the lower half plane that are purely real or purely imaginary on the real axis, the relevant continuations are $f(z^*)^*$ and $-f(z^*)^*$, respectively. Using the fact that $\Phi_{13}(\kappa)$ is real and $\Phi_{23}(\kappa)$ is imaginary on the real axis below the critical value, we can then write 
\begin{equation}
\begin{aligned}
    \Phi_{12}^{\text{up}}(\kappa)&=\Phi_{13}(\kappa^*)^*+\Phi_{23}(\kappa^*)^*\\
    &=\Phi_{12}^{\text{down}}(\kappa^*)^*+2\Phi_{23}(\kappa^*)^*.
\end{aligned}
\end{equation}
The general rule is that all curves may be constructed via sums of integer multiples of other action cycles, determined by the above principles or by continuity. This procedure gets slightly more involved the higher the power of $N$, since there can be additional regions of the plane that have jumps from the representation, but the general principle is the same as outlined above. We give the formulae for all curves required to construct the Stokes eyes for $N=3,4,6$ in the supplementary materials. There we also provide numerical codes to reproduce the locations of the resonance poles (Jost zeros) in each of these cases for the relevant partial waves. The method used is numerical integration along complex-rotated contours in conjunction with the argument principle; these methods were corroborated by semi-analytic techniques outlined in \cite{Gaor3,Gaor6,Birser3,Spectorr4,Jie_2025} that compute the zeros in terms of conditions on continued fractions.

\bibliography{occulus.bib}

\end{document}